\documentstyle[12pt, cite, a4]{article}

\begin{document}
\begin{center}

\bigskip
{\Large\bf 
Multifractal Multiplicity Distribution \\ 
in Bunching-Parameter Analysis}

\vspace{2.0cm}

{\large\rm  S.V.Chekanov\footnote {Present address: 
High Energy Physics Institute Nijmegen (HEFIN), University of 
Nijmegen/NIKHEF, NL-6525 ED Nijmegen, The Netherlands }
and V.I.Kuvshinov}

\bigskip

{\it Institute of Physics, Academy of Sciences of Belarus\\
F.Scaryna Av. 70, 220072 Minsk, Belarus}

\vspace{1.0cm}

{\large\em  Published in J. of Phys. G: Nucl. Part. Phys. V22 (1996) 601-610}

\vspace{1.0cm}

\end{center}

\begin{abstract}
A new multiplicity distribution with multifractal properties
which can be used in high-energy physics and quantum optics is
proposed.  It may be considered as a generalization of 
the negative-binomial distribution.  
We find the structure of  the generating
function for such distribution and discuss its properties.
\end{abstract}

%%%%%%%%%%%%%%%%%%%%%%%%%%%%%%%%%%%%%%%%%%%%%%%%%%%%%
\bigskip
\centerline {\large\bf 1.~Introduction}
\medskip
%%%%%%%%%%%%%%%%%%%%%%%%%%%%%%%%%%%%%%%%%%%%%%%%%%%%%

Multifractal analysis in high-energy physics 
and quantum optics   has received a great
interest in recent years due to the possibility to obtain 
quantitative and qualitative results concerning multiparticle 
production in different processes.
In addition, the analysis is becoming an important 
theoretical tool to
discriminate between different dynamical models.

The fact that the normalized factorial moments (NFMs) 
\begin{equation}
F_{q}=\frac{\langle n(n-1)\ldots
(n-q+1)\rangle }{\langle n\rangle ^q},
\label{79}
\end{equation}
($n$ is the number of particles in a 
restricted phase-space interval $\delta$,
$\langle\ldots\rangle$
is the average over all events) depend on
the size of the phase-space interval (bin)  as
$F_{q}\sim\delta^{-\phi_{q}}$
(intermittency phenomenon) is a manifestation 
of non-statistical fluctuations in the multiplicity 
distribution  of secondary particles
produced in high-energy physics \cite{1,2,3,4}. 
The multifractal behavior, when the anomalous
fractal dimension $d_{q}=\phi_{q}/(q-1)$  depends on $q$, 
is particularly important because such behavior is typical 
for the vast majority of  experiments \cite{2,3,4}.
It is more pronounced in two- and three-dimensional 
phase-space domains. This behavior has also been found 
in photon-counting experiments on laser
fluctuations near threshold, where $\delta$ 
is the counting time interval $T$ \cite{5}.
Thus, the problem of multifractality is a common one 
both for high-energy physics and quantum optics.

In this paper, we shall discuss the 
theoretical aspect of the problem 
in the context of high-energy physics by means of  
bunching parameters (BPs) \cite{6}.
We introduced this quantity in order
to get a simple and efficient method for the analysis  
of complex multiplicity distributions in 
restricted phase-space intervals.
Our consideration  can also be used in any field of
research, where local dynamical fluctuations 
are a subject of investigation.

As is well known, the negative-binomial 
distribution  has become
the focus of interest in view of its 
applicability to the study of
multiparticle productions in high-energy physics. 
However, it has been  noted that this distribution
has no multifractal properties for small 
phase-space intervals \cite{6,7}.

Moreover, the study of 
charged-particle multiplicity distributions
in restricted rapidity intervals conforms that the 
negative-binomial distribution is not
sufficient to describe the data in $Z^0$ hadronic decays due to
a shoulder structure of the experimental 
distributions \cite{8}.
A similar conclusion was 
obtained by ALEPH collaboration \cite{9}:
it was shown that the negative-binomial 
distribution does not 
describe the experimental data,
either in restricted rapidity intervals or in the full phase
space. The UA5 collaboration has also observed that 
the negative-binomial distribution
fails to give a good fit to data at the center-of-mass energy
of 900 GeV \cite{10} in full 
phase space due to a  shoulder structure.
This structure is explained by the 
superposition of 2-jet events with
low multiplicities and 3-jet and 4-jet events yielding much 
larger multiplicities. As a rule, this leads to more complex
multiplicity distribution for full phase space
than the negative-binomial one.

The conclusion must be that the 
negative-binomial distribution is not sufficient to describe the
experimental distributions both for restricted rapidity windows 
and  for full phase space  in
definite experimental situations and, hence,
a true multiplicity distribution must be more complicated.

In this paper we propose a new multiplicity distribution which has
multifractal properties for small phase-space intervals and is very
similar to conventional negative-binomial form for  
large phase-space intervals. 
We shall
analyze this distribution in terms of BPs 
and bunching moments. As we
shall see, such an investigation is  simpler than the analysis  of
multiplicity distributions with the help of NFMs.  
Moreover, recurrence
relations for probabilities can lead to a non-traditional form of
generating functions, both for well-known  distributions 
(Poisson, geometric,
logarithmic, positive-binomial, 
negative-binomial distributions) 
and the multiplicity distribution
with multifractal properties for the small phase-space intervals.

In the second section, we give a short 
introduction to the bunching-parameter method.
In the third section, we consider the general form of a multifractal
negative-binomial like distribution and derive its generating function.
In Sect.~4,  the properties of Markov branching process  leading
to this multiplicity distribution are considered.
In Sect.~5,  we discuss a particular form of such a generalization of 
the negative-binomial distribution
and illustrate its multifractal properties. In Sect.~6, we present
the conclusions.

%%%%%%%%%%%%%%%%%%%%%%%%%%%%%%%%%%%%%%%%%%%%%%%%%%%%%%%%%%%%%%%%
\bigskip
\centerline {\large\bf 2.~Bunching-parameter approach}
\medskip
%%%%%%%%%%%%%%%%%%%%%%%%%%%%%%%%%%%%%%%%%%%%%%%%%%%%%%%%%%%%%%%%

Normalized factorial moments have become an important and popular topic
of experimental and theoretical investigations in 
high-energy physics and quantum optics.
Measuring the NFMs
is equivalent to measuring the multiplicity distribution. 
On the other hand, 
recently another simple mathematical tool to 
investigate the behavior of the multiplicity distribution
in different phase-space intervals has been proposed. 
In order to reveal ``spike'' structure
of events, one can study the behavior of probabilities themselves
by means of the BPs.
The definition of the BPs is given by  the formula \cite{6}
\begin{equation}
\eta_{q}=\frac{q}{q-1}\frac{P_{q}P_{q-2}}
{P_{q-1}^{2}},  \qquad q\ge 2,
\label{1}
\end{equation}
where $P_{q}$ is the probability of finding $q$ particles inside
the limited phase-space interval $\delta$ \footnote{In ref.~\cite{6} we
introduced the BPs by averaging the probabilities over all bins of
equal width.  Here we consider only one fixed bin. We do this only for
the sake of simplicity and it is not a physical restriction.}. 
For example, $\eta_q=1$  for the Poisson distribution. 
If the size of the phase space is small and the average number of particles
in this interval is approximately proportional to $\delta$, then we have the
following approximate relation between the NFMs and the BPs \cite{6}
\begin{equation}
F_{q}\simeq\prod_{i=2}^{q}\eta_{i}^{q+1-i}.
\label{271}
\end{equation}
Such relation, in fact, means that BPs share 
with the NFMs the property of suppression of Poissonian noise.

For the intermittent fluctuations, one  
expects  $\eta_2\propto\delta^{-d_2}$
($d_2$ is the second-order anomalous fractal dimension), 
while the high-order BPs may have
a different dependence on $\delta$.
In \cite{6} it has been shown that for the 
high-energy collisions with the multifractal
behavior of the NFMs $d_{q}=d_2(1-r)+d_2rq/2$, 
the power-like behavior of the BPs
$\eta_{2}\propto\delta^{-d_{2}}$, 
$\eta_{s}=\eta_{2}^{r}$, $s>2$ is
valid. The positive parameter $r$ can be 
interpreted  as a degree of
multifractality (for $r=0$, we have  exactly  
monofractal behavior).  Thus, the
problem of the multifractal multiplicity distribution 
with inverse power $\delta$-dependence
of {\it all} BPs is central one.
 
As discussed in \cite{6}, 
the use of the BPs can give a general answer to the
problem of finding a multiplicity distribution leading to observed
intermittency. 
Indeed, any multiplicity distribution can be expressed as
\begin{equation}
P_n=\frac{P_0}{n!}\left(\frac{P_1}{P_0}\right)^n
\prod_{i=2}^{n}\eta_i^{n+1-i}.
\label{272}
\end{equation}
On the contrary, a multiplicity distribution can not be 
expressed in terms of its factorial
moments if this distribution is not truncated one \cite{4}.  

\newpage

%%%%%%%%%%%%%%%%%%%%%%%%%%%%%%%%%%%%%%%%%%%%%%%%%%%%%%%%%%%%%%%%%%%%
\bigskip
\centerline {\large\bf 3.~Recurrence 
relation for multifractal multiplicity distribution}
\medskip
%%%%%%%%%%%%%%%%%%%%%%%%%%%%%%%%%%%%%%%%%%%%%%%%%%%%%%%%%%%%%%%%%%%%

A necessary and sufficient condition for a 
multiplicity distribution $P_{n}$
to be  negative-binomial one ($P_{n}^{N}$) is the recurrence relation 
\cite{11}
\begin{equation}
\frac{P_{n}^{N}}{P_{n-1}^{N}}=C_{n-1}=
\frac{a+b(n-1)}{n},  \qquad n\ge 1, 
\label{3}
\end{equation}
where $a$, $b$ are positive constants for  fixed $\delta$ ($b<1$).
Note that throughout this paper we shall treat the probabilities
and the parameters $C_{n-1}$, $a$, $b$ as continuous
functions of the phase-space interval of size $\delta$.
If $a\ne 0$,
$b\ne 0$, iterations of (\ref{3}) with the normalization condition
$\sum_{n=0}^{\infty}P_{n}^{N}=1$ give the negative-binomial distribution. 
In the limit
$b\to 0$,
the recurrence relation gives a Poisson distribution.
The case $a\to 0$
at constant $b$ shows that the 
negative-binomial distribution reduces to a 
logarithmic distribution.
For $a=b$, we get a geometric multiplicity distribution.
For $a>0$, $b<0$ and $a/b$ integer, we have a 
positive-binomial distribution. 
Using the definition of the BPs, (\ref{3}) and theorem
of \cite{6}, it is
easy to see that none of  these distributions  leads
to multifractality  if any assumption is  chosen for
the behavior of $a$, $b$ in small phase-space bins.

Thus, it is important to find some multifractal generalization of
the commonly used  multiplicity distributions. 
From a physical point of view, 
to find such a multiplicity distribution, in fact,  means to 
understand the reasons of intermittency with  multifractal behavior
of the anomalous fractal dimensions. 
However, the level of theoretical understanding of this 
phenomenon is still 
insufficient (hadronization problem) 
and is quite different for various types of collision processes \cite{4}.
Nevertheless, from a mathematical point of view, we can propose a
distribution that has {\em a priori} multifractal behavior.

There is a natural way to include  BPs with  power-law behavior
into a new recurrence relation in order to obtain a modification
of (\ref{3}) which can generate the multifractal
multiplicity distribution in the limit of small $\delta$. 
To see this, let us rewrite the definition (\ref{1}) of BPs as
follows
\begin{equation}
\frac{P_{q}}{P_{q-1}}=
\eta_q\frac{q-1}{q}\frac{P_{q-1}}{P_{q-2}}.
\label{4i}
\end{equation}
As mentioned in Sect.~2, for a  multifractal behavior with
$r=1$, all BPs have the same power-law behavior, i.e.,
in the simplest case,
we can write  
\begin{equation}
\eta_q=g\propto \delta^{-\beta}.
\label{4ii}
\end{equation}
The $\beta$
is a positive constant and is taken as a measure of the strength of
the multifractal effect.
After that, by combining (\ref{3}) and (\ref{4i}),
we assume the following recurrence relation
\begin{equation}
\frac{P_{n}}{P_{n-1}}=
\frac{a+b(n-1)}{n}+g\frac{n-1}{n}
\frac{P_{n-1}}{P_{n-2}},  
\label{4}
\end{equation}
where $n\ge1$.
Here, 
in fact, $g$ can be either a new parameter or some combination
of the parameters
$a$ and $b$ (the latter case we shall discuss below). 
Equation (\ref{4}) is a  sufficient condition 
to construct a distribution which
for a large phase-space interval ($g\ll 1$) is very 
similar to the multiplicity distribution
(\ref{3}) and
has the multifractal behavior for a small phase-space interval
(the value of $g$  is not small).

The expression (\ref{4}) can be rewritten as
\begin{equation}
\frac{P_{n}}{P_{n-1}}=\tilde C_{n-1}=
\sum_{l=0}^{n-1}\frac{g^{l}}{n}(a+b(n-l-1)),
\label{5}
\end{equation}
with the sum equal to
\begin{equation}
\left\{ \begin{array}{ll}b(n-ng)^{-1}[k(1-g^n)+n+
(g^n-1)(1-g)^{-1}], & \mbox {for $g\ne 1$}, \\
b(k+0.5(n-1)), & \mbox {for $g=1$,}
\end{array}
\right.
\label{55}
\end{equation}
where $k^{-1}=b/a$ was called aggregation coefficient by
Giovannini and Van Hove \cite{11} 
for the usual negative-binomial distribution.
For a compact description of the recurrence relation,
we shall use expression for $\tilde C_{n-1}$
in the form of sum (\ref{5}).
Using (\ref{4}) and (\ref{5}), we obtain for BPs 
\begin{equation}
\eta_{q}=
\frac{1+k^{-1}(q-1)}{\sum_{l=0}^{q-2}g^{l}(1+k^{-1}(q-l-2))}+g.
\label{6}
\end{equation}
We have multifractal behavior  because the second
term in (\ref{6}) has power-law behavior and, 
for enough  small $\delta$,
can  be the leading one. The multiplicity distribution
corresponding to (\ref{4}), (\ref{5}) is
\begin{equation}
P_{n}=\frac{P_{0}}{n!}\prod_{s=0}^{n-1}\sum_{l=0}^{s}
g^{l}(a+b(s-l)).
\label{7}
\end{equation}
Throughout this paper, we consider the probability
$P_{0}=1-\sum_{n=1}^{\infty }P_{n}$ of having no 
particles in $\delta$ as a
normalization constant. 
Since in the limit $g\to 0$ expression (\ref{7})
reduces to a negative-binomial distribution, we shall call 
this distribution a   
multifractal negative-binomial distribution (MNBD) 
and denote it  as $P^{M}_{n}$.

Let us note that we can analyze a multiplicity distribution  
written in terms of
the recurrence relations using the generating function 
for $P_{n}/P_{n-1}$. Let 
\begin{equation}
G(z)=
\sum_{n=1}^{\infty}\frac{z^{n}}{(n-1)!}\frac{P_{n}}{P_{n-1}}.
\label{8}
\end{equation}
be the generating function for the ratios $P_{n}/P_{n-1}$.
Then, the BPs are given by
\begin{equation}
\eta_{q}=
\frac{G^{(q)}(z)\mid_{z=0}}{G^{(q-1)}(z)\mid_{z=0}}.
\label{9}
\end{equation}
For example, the generating function of 
the negative-binomial distribution has 
the following form
\begin{equation}
G^{N}(z)=(a-b)(e^{z}-1)+b\>z\>e^{z}.
\label{10}
\end{equation}
For the Poisson distribution, we have
\begin{equation}
G^{P}(z)=a(e^{z}-1).
\label{11}
\end{equation}
In terms of the generating function, 
the recurrence relation (\ref{4})
can be rewritten in form of the 
following differential equation
\begin{equation}
\frac{dG^{M}(z)}{dz}=\frac{dG^{N}(z)}{dz}+g\>G^{M}(z),
\label{12}
\end{equation}
with the initial condition (see (\ref{8}))
\begin{equation}
G^{M}(z=0)=G^{N}(z=0)=0.
\label{13}
\end{equation}
Using (\ref{12}), 
we can obtain 
the generating function in integral form
as follows
\begin{equation}
G^{M}(z)=G^{N}(z)+\sum_{i=1}^{\infty}g^{i}\int G^{N}(z)d^{i}z.
\label{213}
\end{equation}
Using (\ref{10}) and condition (\ref{13}), 
one gets the analytical solution of the equation (\ref{12})
\begin{equation}
G^{M}(z)=\frac{e^z}{1-g}\left(a+b\>z-\frac{b}{1-g}\right)-e^{gz}
\frac{a(1-g)-b}{(1-g)^2},
\label{14}
\end{equation}
for $g\ne 1$ and, using (\ref{5}) and (\ref{8}),  
$G^M(z)=e^z(az+bz^2/2)$ for $g=1$. Below,  we
shall not consider the trivial case, when $g=1$.

\medskip
Now let us mention two limiting cases:

a) In the simplest Poisson case, when $b=0$, we have the following
multifractal Poisson distribution with the 
generating function of the form
\begin{equation}
G^{MP}(z)=\frac{a}{1-g}\left(e^z-e^{zg}\right).
\label{15}
\end{equation}
Then, from (\ref{6}) and (\ref{7}), one has
\begin{equation}
P^{MP}_{n}=P_{0}\frac{a^{n}}{n!}\prod_{s=0}^{n-1}\sum_{l=0}^{s}g^l,
\label{16}
\end{equation}
\begin{equation}
\eta_{q}^{MP}=\frac{1}{\sum_{l=0}^{q-2}g^l}+g.
\label{17}
\end{equation}
Here $a$ is not an average multiplicity as 
for a usual Poisson distribution.

b) The limit  $a\to 0$ is 
also interesting since it leads to
the multifractal logarithmic distribution 
\begin{equation}
G^{ML}(z)=\frac{be^z}{1-g}
\left(z-\frac{1}{1-g}\right)+e^{gz}\frac{b}{(1-g)^2}.
\label{18}
\end{equation}
From (\ref{6}) and (\ref{7}), one gets
\begin{equation}
P^{ML}_{n}=P_{1}\frac{b^{n-1}}{n!}
\prod_{s=0}^{n-2}\sum_{l=0}^{s}g^{l}(s-l+1),
\label{19}
\end{equation}
\begin{equation}
\eta_{q}^{ML}=\frac{q-1}{\sum_{l=0}^{q-2}g^l(q-l-2)}+g,
\label{20}
\end{equation}
where $q>2$ and $P_{1}$ is a normalization constant (for 
the logarithmic multiplicity distribution, $P_0=0$).
We see that these multiplicity 
distributions have the same power law
behavior of the high-order BPs for small $\delta$. 
In this sense, the
distributions are equivalent for small phase-space intervals.

\medskip

It is important to emphasize here that an infinite sequence
of probabilities $P_n$  can be normalized if it 
converges, i.e., if $P_n\to 0$ for $n\to\infty$.
This is possible if, 
for every $i$ greater than some number $\zeta$,
we have the following condition
\begin{equation}
\frac{P_i}{P_{i-1}}<1
\label{220}
\end{equation}
(ratio test). For the MNBD (\ref{5}) 
this is possible if
\begin{equation}
0<g<1, \qquad 0<\frac{b}{1-g}<1.
\label{221}
\end{equation}
For any  other domain of $g$, 
we must truncate the MNBD, putting $P_i^M=0$
for $i>\zeta$.

One can understand, from  definition (\ref{1}), that BPs are
sensitive to the local multiplicity fluctuations 
(or to the behavior of the multiplicity distribution in  
small phase-space intervals)
near the multiplicity  $q=n-1$. 
In order to study the  total contribution from 
multiplicity  fluctuations for  large values  of $n$, it is appropriate
to introduce the ``bunching moments'' $b_q$ as follows
\begin{equation}
b_q\equiv G^{(q)}(z)\mid_{z=1}=
\sum_{s=q}^{\infty}\frac{s}{(s-q)!}\frac{P_{s}}{P_{s-1}},
\label{88}
\end{equation}
following an analogy with  factorial moments. 
The knowledge of the generating function
gives us a possibility to calculate 
both the BPs and the bunching moments.
The higher the rank of the $b_q$, 
the more sensitive they are to the
``tail'' of the multiplicity distributions for large $n$. 
In order to  normalize  the bunching moments $B_{q}$,
we put
\begin{equation}
B_q\equiv \frac{b_q}{b_1}=\frac{G^{(q)}\mid_{z=1}}{G^{'}\mid_{z=1}}.
\label{89}
\end{equation}
Then for the Poisson distribution (\ref{11}) we have
\begin{equation}
B^P_q=1,  \qquad q=1\ldots\infty.
\label{90}
\end{equation}
For the negative-binomial distribution (\ref{10}):
\begin{equation}
B^N_q=\frac{k+q}{k+1}, \qquad q=1\ldots\infty.
\label{91}
\end{equation}
This  means that the negative-binomial distribution 
is broader than the Poisson distribution 
($B_q^N>B_q^P$).
For the geometric multiplicity distribution ($k=1$), 
the normalized bunching moments  are
larger than those of the 
negative-binomial distribution with $k>1$. In this case, we can say
that the geometric  distribution is broader 
than the negative-binomial one.
For the positive-binomial multiplicity
distribution, where   $k<0$ and  integer, the normalized
bunching moments are smaller than unity 
because this  distribution is narrower than
Poisson. 

For MNBD (\ref{14}),  one gets
\begin{equation}
B^M_q=\frac{k(g-1)(g^qe^{g-1}-1)+g^qe^{g-1}-g(q+1)+q}
{k(g-1)(ge^{g-1}-1)+ge^{g-1}-2g+1}, 
\label{92}
\end{equation}
$q=1\ldots\infty$.
For  a  small $g$ (or for a large
phase-space interval), the MNBD slightly differs 
from the negative-binomial distribution.
For $g\to 0$ ($B^M_q\to B^N_q$), the MNBD tends to 
the negative-binomial distribution.
The situation drastically changes when   
large $g$ (or  small $\delta$)
is considered. 
The $B^M_q$ becomes larger than $B^N_q$, since the MNBD
becomes {\em broader} than the 
negative-binomial distribution for all $k$. 
Bear in mind of this property, it is
not surprising that for this region of phase space the 
MNBD reflects the
increase of the local intermittent 
fluctuations and can lead to a  multifractal
behavior of the anomalous fractal dimension.

The generating function 
$Q=\sum_{n=0}^{\infty}z^{n}P_{n}$ commonly
considered in probability theory and its 
applications can also be used in
the multifractal generalization \cite{12}. 
To analyze this generating
function, it is appropriate to use the usual NFMs or 
the normalized cumulant moments \cite{4}.
In fact, these two "languages", using the BPs and the NFMs 
for the study of multiplicity
distributions in terms of the 
generating functions $G$ and $Q$, are equivalent.
Nevertheless, in some cases, the analysis of multiplicity 
distributions in small $\delta$
with the help of BPs and the bunching moments is 
simpler, because an analytical
form of the NFMs for some multiplicity distributions
may be too complicated.
For example, 
to find a simple form of the generating functions for  
multifractal distributions (\ref{7}), (\ref{16}), (\ref{19}), 
is rather difficult. 
The technical advantage of the use of these quantities
comes mainly from the fact that, as a rule, the structure of the ratio
$P_n/P_{n-1}$ is simpler than the form of the probabilities
$P_n$ themselves.

It is appropriate to make some 
remarks here on the relationship between
the form of the recurrence relation (\ref{4}) and the 
definition of the generating function (\ref{8}).
The particular form of recurrence relation (\ref{4})
was chosen  because of
its simplicity. 
For example, the factor $(n-1)/n$ in the last term
of (\ref{4}) was chosen only 
because it is possible to rewrite this relation
in the form of generating function (\ref{8}).
It is easy to check that other similar forms involving
$P_{n-1}$ and $P_{n-2}$ in the 
recurrence relation can lead to qualitatively
similar results, providing  a  singular form of BPs for all order.
However, in these cases, one needs to introduce 
other forms of  generating function in
order to obtain  a differential equation 
for the MNBD with simplest solutions.

%%%%%%%%%%%%%%%%%%%%%%%%%%%%%%%%%%%%%%%%%%%%%%%%%%%%%%%%%%%%%
\bigskip
\centerline {\large\bf 4.~The MNBD as a solution of 
non-linear Markov process}
\medskip
%%%%%%%%%%%%%%%%%%%%%%%%%%%%%%%%%%%%%%%%%%%%%%%%%%%%%%%%%%%%%%

Now we shall show that the form of the  MNBD can be obtained
from a stochastic Markov process with non-linear birth rate.
Let $P_n(t)$ be the probability to have $n$ particles at time
$t$. Of course, such a choice of an 
evolution parameter is not unique.
In principle, the evolution variable $t$ can be connected, 
for example, with the squared mass of the branching parton in
the parton shower. For simplicity, we shall assume that the
process starts at time $t=0$, with the initial condition
$P_0(t=0)=1$, $P_n(t=0)=0$, $n>0$.
We shall consider a very general  
birth-death process with an infinitesimal birth rate $w_n^+$
and an infinitesimal death rate $w_n^-$ of particles, treating these 
parameters  as continuous functions of $t$. The corresponding Markov
equation is
\begin{equation}
P'_n(t)=
w_{n-1}^{+}P_{n-1}(t)+w_{n+1}^{-}P_{n+1}(t)-(w_n^++w_n^-)P_n(t),
\label{111}
\end{equation}
\cite{13}. For a stationary process, 
when time goes to infinity, the
$P_n(t)$ are  $t$-independent constants.  
Then, from (\ref{111}) one has
$\pi_n-\pi_{n+1}=0$, $n=1,2,..$, 
$\pi_n=w_n^-P_n-w_{n-1}^+P_{n-1}$. Taking into account $\pi_0=0$,
we have $\pi_n=0$, $n\geq 1$ and, hence, a  solution of (\ref{111}) 
in the form of the recurrence relation 
\begin{equation}
\frac{P_n}{P_{n-1}}=\frac{w_{n-1}^+}{w_n^-}.
\label{112}
\end{equation}
The negative-binomial distribution can be 
considered as a  stationary solution of (\ref{111})
if we put the linear forms for $w_n^+$ and $w_n^-$,
\begin{equation}
w_n^+=\gamma +\beta n, \qquad w_n^-=\rho n,
\label{113}
\end{equation}
where $a=\gamma /\rho$ and $b=\beta /\rho$ ($\gamma$, $\beta$, $\rho$
are $t$-independent ). If we admit that the parameter $w_n^+$ is
the non-linear function of $n$, i.e.,
\begin{equation}
w_n^+=\sum _{l=0}^{n}g^l[\gamma +\beta (n-l)], \qquad w_n^-=\rho n,
\label{114}
\end{equation}
then the MNBD can be found as a stationary solution of the evolution
equation (\ref{111}). Here, in fact, parameter $g$ represents the strength
of influence of the non-linearity in the equation. For $\delta\to 0$
the non-linearity  of the birth rate increases.

So, the multifractal structure of MNBD can be caused by the 
non-linearity of the stationary Markov process.
From the point of view of high-energy physics, 
the multiparticle production  in QCD and the
subsequent transition to hadrons have a strongly non-linear nature.
In this sense, the form of (\ref{114}) may reflect  the
non-linearity of the underlying multiparticle dynamics leading to
multifractality.

\newpage

%%%%%%%%%%%%%%%%%%%%%%%%%%%%%%%%%%%%%%%%%%%%%%%%%%%%%%%%%%%%%%%%%
\bigskip
\centerline{\large\bf 5.~The choice of 
parameter $g$ and application}
\centerline{\large\bf to experimental data}
\medskip
%%%%%%%%%%%%%%%%%%%%%%%%%%%%%%%%%%%%%%%%%%%%%%%%%%%%%%%%%%%%%%%

Up to now, we have considered $g$ as a free parameter with
the power-law behavior $g\propto\delta^{-\beta }$. 
The next question is how
does one choose  $g$, if one  wants  
to obtain the multiplicity 
distribution with a degree of
multifractality $r$ (see Sect.~2).
The simplest way is to assume the following form 
\begin{equation}
g=\omega (r)\left(\frac{b}{a}\right)^{r}=\omega (r)k^{-r},
\label{24}
\end{equation}
where $\omega (r)$ is some function  which tends to zero for $r\to 0$
(in the simplest case, $\omega (r)=r$).
The  parameter  $r$ allows interpolation between the 
negative-binomial distribution with  monofractal
behavior ($r=0$) and the MNBD with the multifractal behavior ($r>0$).
Let us remind
that the aggregation coefficient $k^{-1}$
is related to the mean multiplicity $<n>=a/(1-b)$
and the dispersion $D$ of the negative-binomial distribution as
$D^{2}=<n>+<n>^{2}k^{-1}$.

\medskip
From the point of view of high-energy physics, 
our choice is justified  by the following reasons:

i) If we have the negative-binomial distribution, 
for a large (pseudo)rapidity  interval  $k^{-1}$ has
a small value ($k^{-1}\sim 0.1-0.01$). Applying a 
fit by (\ref{7}), we can
hope that $g$ in the MNBD will be small also 
and this  distribution will be
similar to the negative-binomial distribution.

ii) It is essential that the negative-binomial distribution 
is approximately valid,  not only for  large
phase-space intervals,  but also for  small ones.
For the negative-binomial distribution, the behavior 
$k^{-1}\propto \delta^{-\beta }$ lies
in the framework of the assumption that intermittency is governed
only by the aggregation coefficient. In \cite{7} Van Hove showed
that, if $k^{-1}\propto \delta^{-\beta }$ for small $\delta $, 
then the negative-binomial distribution
has  monofractal behavior, $F_q\propto\delta^{-d_q(n-1)}$,
$d_q=\beta $. Then $\eta_{2}\propto \delta^{-\beta}$,
$\eta_{s}=const.$, $s>2$ \cite{6}. In our case, 
the assumption $\omega (r)\ne 0$ 
yields intermittency with the multifractal behavior of 
the anomalous fractal dimension.

From the experimental point of view, a
good approximation for the aggregation
coefficient of the negative-binomial distribution is \cite{14}
\begin{equation}
k^{-1}=c\>M^{d_{2}}, \qquad M=\frac{\Delta}{\delta},
\label{26}
\end{equation}
where $\Delta$ and $\delta$ are
three-dimensional full and 
limited phase-space intervals, respectively,
$d_{2}$ is the anomalous fractal dimension  of the second order.
For different reactions, 
the parameters $c$ and $d_{2}$ are of the same
order of magnitude, moreover, the $d_{2}$ has almost universal value
$\sim 0.4$ \cite{14}. Expression (\ref{26}) has been  
obtained from the well-known
relation $F_{2}^{N}=1+k^{-1}$ between second-order NFM and the
aggregation coefficient which is correct for the 
negative-binomial distribution.

iii)  Since the fits for different reactions show a
logarithmic increase of $k^{-1}$ with energy $\sqrt{s}$ in
the negative-binomial distribution, 
one may expect that for the MNBD we shall obtain
the similar effect. Then, the
distribution (\ref{7}) in the full phase space may 
slightly differ from the standard negative-binomial distribution 
for large energies.
This is quite important because, as mentioned already in the
introduction, the usual negative-binomial distribution fails to
describe multiplicity distributions at
900GeV in $p\bar p$ collisions and in $Z^{0}$ hadronic decay
for  full phase space.

\medskip

Let us obtain the anomalous fractal dimension  for the MNBD 
in the particular case, when $g=\omega (r)k^{-r}$.
The calculation of the NFMs can be simplified when an average
multiplicity in $\delta$ is small. Then, the NFMs are given by 
expression (\ref{271}). 
Let us discuss two domains of the parameter $r$:

(i) $0\leq r\leq 1$:  From (\ref{6})
we have
\begin{equation}
\eta_2=1+k^{-1}+w(r)k^{-r}, 
\label{288}
\end{equation}
\begin{equation}
\eta_{s}=\frac{1+k^{-1}(s-1)}{\sum_{l=0}^{s-2}\omega^{l}(r)k^{-rl}
(1+k^{-1}(s-l-2))}+w(r)k^{-r},  \quad s>2.
\label{289}
\end{equation}
We see that, if $k\propto \delta^{d_2}$, 
the leading terms of the BPs have the
following behavior: $\eta_{2}\propto\delta^{-d_{2}}$,
$\eta_{s}\propto\delta^{-rd_{2}}$, $s>2$. Then
\begin{equation}
F_{q}\propto\delta^{-d_{q}(q-1)}
\qquad d_{q}=d_{2}(1-r)+d_{2}r\frac{q}{2}.
\label{28}
\end{equation}
For $r=0$, we have the monofractal behavior $d_{q}=d_{2}$, and the MNBD
reduces to the negative-binomial distribution.

(ii) $r>1$:  The leading terms of the BPs are given by the expression
$\eta_{q}\propto\delta^{-rd_{2}}$, $q\ge 2$. The corresponding 
anomalous fractal dimension  is
\begin{equation}
d_{q}=d_{2}r\frac{q}{2}.
\label{29}
\end{equation}
The values of $r$ for different 
reactions have been discussed in \cite{6}.

\bigskip
\centerline {\large\bf 6.~Conclusion}
\medskip

We have proposed a new multiplicity distribution with  multifractal
properties for small phase-space intervals, basing on the simplicity of
the analysis of multifractality in  terms of BPs and bunching
moments. Guided by the fact that, for the multifractality of
normalized factorial moments the same power-law behavior for all orders
of the BPs is necessary and sufficient, in this paper we focus our
attention on the analysis of the MNBD with the
multifractal behavior.  The MNBD may be considered as a generalization
of the negative-binomial distribution with a new free parameter 
$g$ (or $r$ for the particular case,
when $g=\omega(r)k^{-r}$) which has a power-like behavior for  
small phase-space intervals.

Theoretically, the question 
arises what is the physical reason of such
multifractal distribution (or similar to it).
The problem of multifractality is very complicated and requires
further examination  both in high-energy physics and quantum optics.
Note, as an example, that some versions of cascade
models can lead to this distribution because they have  the same 
anomalous fractal dimension 
(see $\alpha$-model \cite{15}, where the anomalous fractal dimension  
has the form
(\ref{29}) with $r=1$).

On the experimental side of the question, we hope that the MNBD
is interesting since it yields a new possibility to describe
experimental data.

\bigskip
{\large\bf Acknowledgements}
\medskip

This work is part of the research program of the ``Stichting voor
Fundamenteel Onderzoek der Materie (FOM)'', which
is financially supported by the ``Nederlandse Organisatie voor
Wetenschappelijk Onderzoek (NWO)''.

The research of S.V.C. was supported in part
by Grant MP-19 of the Fund for Fundamental Research
of the Republic of Belarus. S.V.C would like to 
thank the High Energy
Physics Institute Nijmegen (HEFIN) for the hospitality. 

{}
\end{document}